\documentclass[12pt]{article}
\usepackage{graphicx}

\newcommand\pubnumber{DPF2015-454}
\newcommand\pubdate{\today}

\def\napoli{Department of Physics, University of Michigan\\
Ann Arbor, MI, 48109, USA}
\def\support{\footnote{Work supported by the United States Department of Energy Early Career Grant under contract DE-SC0008062.}}

\def\Title#1{\begin{center} {\Large #1 } \end{center}}
\def\Author#1{\begin{center}{ \sc #1} \end{center}}
\def\Address#1{\begin{center}{ \it #1} \end{center}}

\newcommand\pubblock{\rightline{\begin{tabular}{l} \pubnumber\\
         \pubdate  \end{tabular}}}
\newenvironment{Abstract}{\begin{quotation}  }{\end{quotation}}
\newenvironment{Presented}{\begin{quotation} \begin{center} 
             PRESENTED AT\end{center}\bigskip 
      \begin{center}\begin{large}}{\end{large}\end{center} \end{quotation}}
\def\Acknowledgments{\bigskip  \bigskip \begin{center} \begin{large}
             \bf ACKNOWLEDGMENTS \end{large}\end{center}}

\def\beq{\begin{equation}}
\def\eeq#1{\label{#1}\end{equation}}
\def\eeqn{\end{equation}}

\def\beqa{\begin{eqnarray}}
\def\eeqa#1{\label{#1}\end{eqnarray}}
\def\eeqan{\end{eqnarray}}

\let\bar=\overbar

\def\Dslash{\not{\hbox{\kern-4pt $D$}}}
\def\dslash{\not{\hbox{\kern-2pt $\del$}}}

\def\msb{{\bar{\ssstyle M \kern -1pt S}}}

\begin{document}
\begin{titlepage}
\pubblock

\vfill
\Title{Electroweak Results from the ATLAS and CMS Experiments}
\vfill
\Author{Junjie Zhu\support}
\Address{\napoli}
\vfill
\begin{Abstract}
I summarize an extensive ATLAS and CMS electroweak physics program that involves a variety of single boson, diboson, triboson, and vector boson scattering measurements. The relevance of these studies 
to our understanding of the electroweak sector and electroweak symmetry breaking is emphasized. I describe the recent results and prospects for future measurements.
\end{Abstract}
\vfill
\begin{Presented}
DPF 2015\\
The Meeting of the American Physical Society\\
Division of Particles and Fields\\
Ann Arbor, Michigan, August 4--8, 2015\\
\end{Presented}
\vfill
\end{titlepage}
\def\thefootnote{\fnsymbol{footnote}}
\setcounter{footnote}{0}

\section{Introduction}

Except the gluons, all fundamental particles we know experience electroweak interactions. 
Due to gauge symmetry and chiral symmetry, all particles are required to be massless in 
the Standard Model (SM) theory. SM explains the origin of mass  
by introducing a scalar Higgs field ($\phi$) with a Mexican-hat potential: $V(\phi)=\mu^2 \phi^2 + \lambda \phi^4$ with $\mu^2<0$~\cite{Higgs}. 
The minima of the potential happen for non-zero vacuum expectation values and break the electroweak symmetry.  
Massless Goldstone bosons generated from the symmetry breaking are absorbed by massless gauge bosons 
to become their longitudinal components of the now-massive $W$ and $Z$ bosons. 
Fermions gain their masses via Yukawa couplings with the Higgs field.  
The quantum excitation of the Higgs field is the Higgs boson.

Besides the neutral current and charged current interactions, gauge bosons also have self-interactions 
due to the non-Abelian structure of the $SU(2)_L \times U(1)_Y$ gauge symmetry group in the electroweak sector of the SM~\cite{sm}. 
The allowed triple gauge coupling (TGC) vertices are $WW\gamma$ annd $WWZ$, and the allowed quartic gauge coupling (QGC) vertices 
are $WWWW$, $WWZZ$, $WWZ\gamma$ and $WW\gamma\gamma$. 
The Higgs boson couples to all massive gauge bosons and fermions, it also couples to itself that results in 
$HHH$ and $HHHH$ coupling vertices.

\section{Global Electroweak fits and SM consistency tests}
The electroweak theory is an extremely successful theory with only a few free parameters: $g$, $g’$, $\mu$ and $\lambda$, 
where $g$ and $g’$ are the weak and electromagnetic couplings, $\mu$ and $\lambda$ are the two parameters for 
the Higgs potential. Other parameters likes fermion masses and strong coupling constant enter by radiative corrections. 
With these parameters known, we can make predictions for all other quantities in the electroweak sector. 
In reality we often choose a few quantities measured with high precision to determine these parameters and 
then make predictions for other variables. We have successfully predicted the preferred top and Higgs mass ranges 
before the actual discoveries. With many experimental measurements performed so far, 
we can check the consistency of the electroweak theory and search for new physics beyond the SM. 

\begin{figure}[htb]
\centering
\includegraphics[height=4.in]{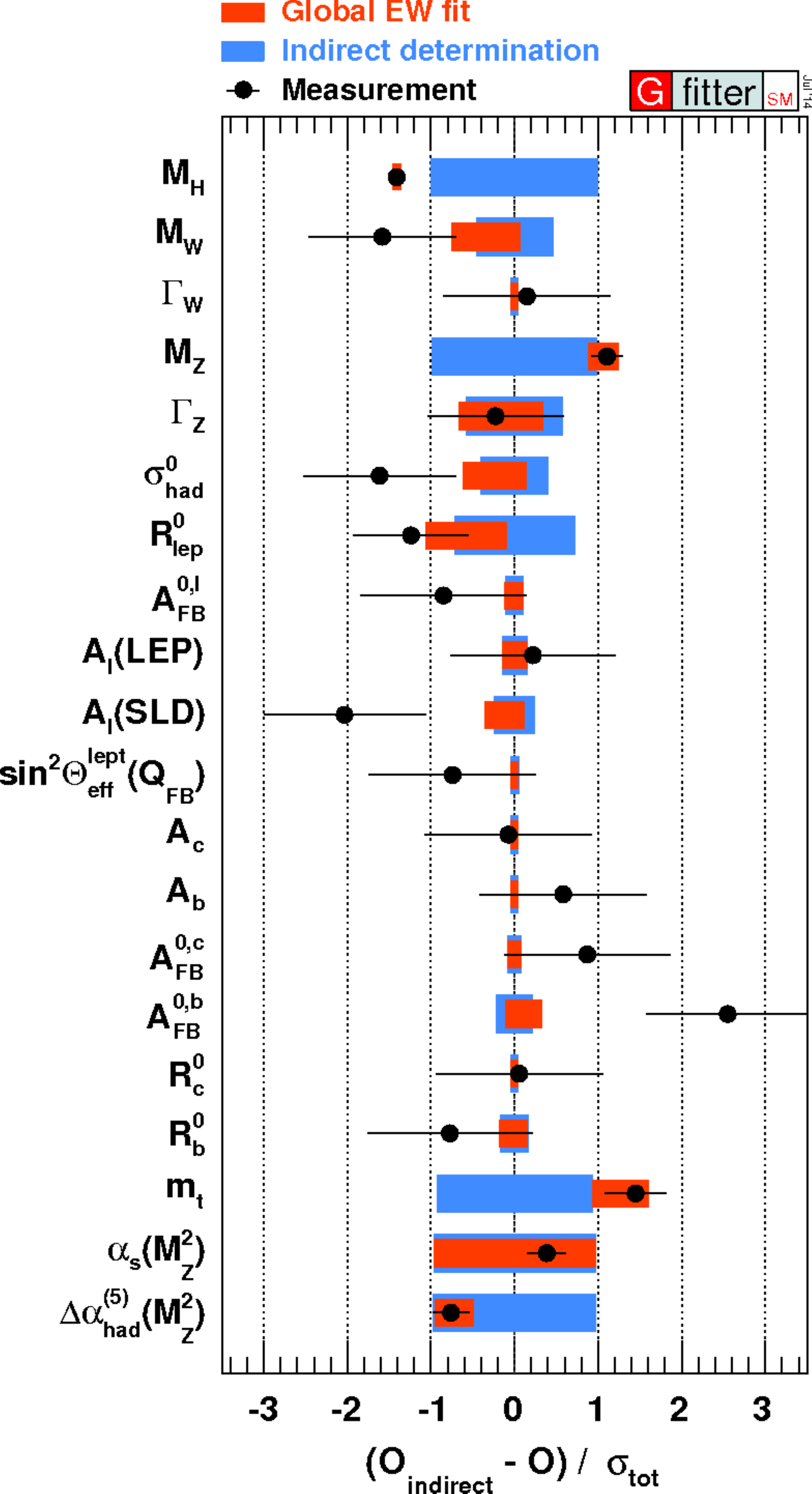}
\caption{Comparing fit results (orange bars) with indirect determinations (blue bars) and direct measurements (data points): pull values for the SM fit defined as deviations to the indirect determinations from the GFitter Collaboration.}
\label{fig:ewk_fits}
\end{figure}

The most recent global electroweak fits was performed by the Gfitter collaboration~\cite{gfitter}. Six measured quantities ($M_Z$, $M_H$, 
$\alpha$, $m_t$, $m_b$ and $m_c$) are used as inputs. 
For theoretical predictions, calculations detailed in \cite{gfitter_theory} and references therein are used. 
The predictions are compared to experimental results from low energy experiments, LEP/SLD, Tevatron and LHC~\cite{exp_results}. 
Figure~\ref{fig:ewk_fits} shows comparison of the global fit results with the
indirect determination in units of the total uncertainty, defined as the uncertainty of the direct measurement
and that of the indirect determination added in quadrature. The indirect determination of an observable
corresponds to a fit without using the corresponding direct constraint from the measurement. The overall $\chi^2 / dof$ is 17.8/14, 
corresponding to a probability of 21\%. There are no individual deviations that exceed 3 $\sigma$, showing good consistency 
between all measurements and theoretical predictions. 

The LHC experiments provided the first direct measurement of the Higgs boson mass ($m_H$)~\cite{mH} which is a critical input to the global fits. 
The uncertainties on the indirect measurement of $m_W$ ($W$ boson mass), $\sin^2 \theta^{\ell}_{eff}$ (lepton effective weak mixing angle), 
$m_t$ and $m_H$ have reduced significantly after the inclusion of the direct $m_H$ measurement in the global fits. 
The fitted $m_W$ has an uncertainty of 8 MeV while the measured value has an uncertainty of 15 MeV. 
The fitted $\sin^2 \theta^{\ell}_{eff}$ has an uncertainty of $7 \times 10^{-5}$, while the measured uncertainty is 
$17 \times 10^{-5}$~\cite{gfitter}. It is thus important to further improve these two measurements. 

\section{$\sin^2 \theta^{\ell}_{eff}$ and $m_W$ measurements}
For the $q\bar{q} \rightarrow X \rightarrow \ell^+ \ell^-$ process, we can have the exchange of 
either a photon or a $Z$ boson. The couplings of the photon to fermions are pure vector couplings, 
while the couplings of the $Z$ boson to fermions are a mixture of vector and axial-vector couplings. 
Due to the presence of both vector and axial-vector couplings, we expect to have an asymmetric distribution of the 
angle between the negatively-charged lepton relative to the incoming quark direction in the dilepton rest frame. 
 Both ATLAS and CMS collaborations have measured this asymmetry using $pp \rightarrow Z/\gamma^* \rightarrow \ell^+ \ell^-$ ($\ell=e, \mu$) events~\cite{atlas_afb}~\cite{cms_afb}. 
Since the LHC is a proton-proton collider, both analyses assumed the longitudinal direction of flight of the $Z$ 
boson is the same direction as that of the incoming quark. The measured asymmetry distributions are found to be consistent with theoretical 
predictions in the whole mass region. ATLAS also extracted $\sin^2 \theta^{\ell}_{eff}$ using events around the $Z$ mass region 
and obtained $\sin^2 \theta^{\ell}_{eff}=0.2308 \pm 0.0005$ (stat) $\pm 0.0006$ (syst) $\pm 0.0009$ (PDF)~\cite{atlas_afb}. 

The current world average uncertainty on $\sin^2 \theta^{\ell}_{eff}$ is $17 \times 10^{-5}$ and is dominated by measurements from lepton colliders~\cite{lep_afb}.  
Due to large production cross section and integrated luminosity expected at the LHC, the statistical uncertainty on $\sin^2 \theta^{\ell}_{eff}$ 
is expected to be negligible but we need to find ways to reduce the dominant PDF uncertainty.
The current best hadron collider measurement comes from D0 with an uncertainty of $47 \times 10^{-5}$~\cite{d0_afb} that is comparable to each individual lepton collider measurement. 
D0 found that using $Z\rightarrow ee$ events in all categories (CC/CE/EE where C means electrons in the central calorimeter and E means electrons in the endcap calorimeter) 
have significantly reduced the $\sin^2 \theta^{\ell}_{eff}$ PDF uncertainty. It is important for the LHC experiments to also include leptons in 
the whole pseudorapidity coverage in their analyses. The targeted overall uncertainty with a total integrated luminosity of 3000 fb$^{-1}$ at 14 TeV is $21 \times 10^{-5}$ for each LHC experiment~\cite{afb_expectation}. 

The current world average uncertainty on $m_W$ is 200 ppm~\cite{tevatron_mW} and is dominated by Tevatron measurements. 
The current best measurement was performed by the CDF 
experiment where the contributions from experimental systematic, theoretical modelling, PDF 
and statistical uncertainties are all close to around 10 MeV~\cite{cdf_mW}. The targeted LHC uncertainty with an integrated luminosity of 3000 fb$^{-1}$ at 14 TeV is 
5 MeV, which requires the PDF and theoretical modelling uncertainty to be around 3 MeV~\cite{afb_expectation}.

Due to large pileup expected at the LHC, the best precision on the $m_W$ measurement might be achieved by fitting the lepton $p_T$ distribution since it 
is less sensitive to detector effects. 
ATLAS studied both PDF and theoretical modelling uncertainties for the 7 TeV $m_W$ measurement~\cite{atlas_mW}. Four different PDF sets are investigated and the 
uncertainty on $m_W$ within a given PDF set is found to be $10-20$ MeV. In addition $10-50$ MeV shifts are found between central values of 
different PDF sets. ATLAS varied the $W$ $p_T$ modelling parameters in {\sc pythia} and found the systematic uncertainty on $m_W$ is close to 7 MeV.

There are many measurements at the LHC that can help reduce PDF uncertainties. One example is the $W$ boson charge 
asymmetry measurement. CMS performed this measurement using 8 TeV data and the measured asymmetry is compared to predictions from three different PDF sets~\cite{cms_WAsy}. 
Experimental uncertainties on all lepton pseudorapidity bins are found to be smaller than PDF uncertainties. 
CMS included this result in the PDF fit and found uncertainties on $u$ and $d$ valence quark PDF distributions have reduced significantly. 
CMS also measured the $W$ and $Z$ inclusive cross sections at 8 TeV~\cite{cms_wzxsection}. Different PDF sets predict different values of production cross sections 
and the ratio. The measured $W$ and $Z$ cross sections and their ratio are found to be consistent with theoretical predictions using different 
PDF sets. 

Both ATLAS and CMS measured the $W$ and $Z$ boson transverse momentum distributions~\cite{atlas_ZpT}~\cite{cms_ZpT}~\cite{cms_WpT}. 
At low boson $p_T$ ($<\sim30$ GeV), the production is dominated by multiple soft gluon emission and 
the resummation technique is needed for theoretical predictions. While at high boson $p_T$ ($>\sim 50$ GeV), the production is dominated by a single hard parton 
emission and perturbative QCD calculation is needed. The measured differential cross sections ($d\sigma/dp_T$) are 
compared to NNLO calculations using {\sc fewz}~\cite{fewz} and predictions from a few event generators. 
The conclusion is that the inclusive cross section agrees with the NNLO calculation, however some shape differences are 
observed between measurements and predictions. These measurements will help us gain a better understanding of the $W/Z$ 
boson production mechanism and further reduce the uncertainty on the $m_W$ measurement.

\section{Processes with Triple Gauge Couplings}
TGC studies can be performed either using diboson processes or using mono-boson production via vector boson scattering (VBS). 
$ZZ\gamma$, $Z\gamma\gamma$ and $\gamma\gamma\gamma$ are forbidden. 
Studies of TGC processes provide fundamental tests of the electroweak sector. New physics 
beyond the SM could introduce anomalous TGCs that can be measured experimentally. A model-independent effective Lagrangian approach 
is often used to study these aTGC parameters. 

\begin{figure}[htb]
\centering
\includegraphics[height=2.2in]{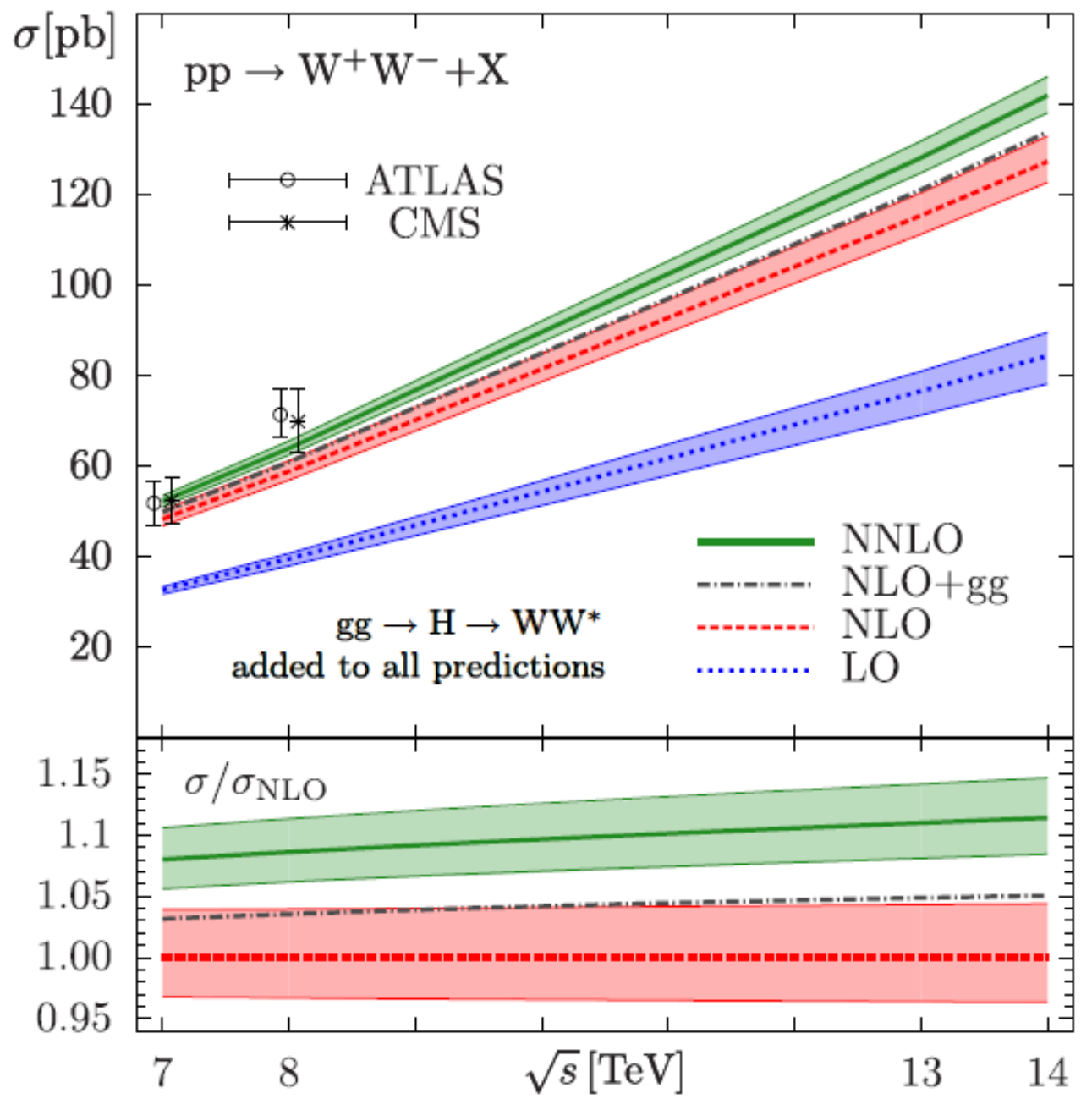}
\includegraphics[height=2.3in]{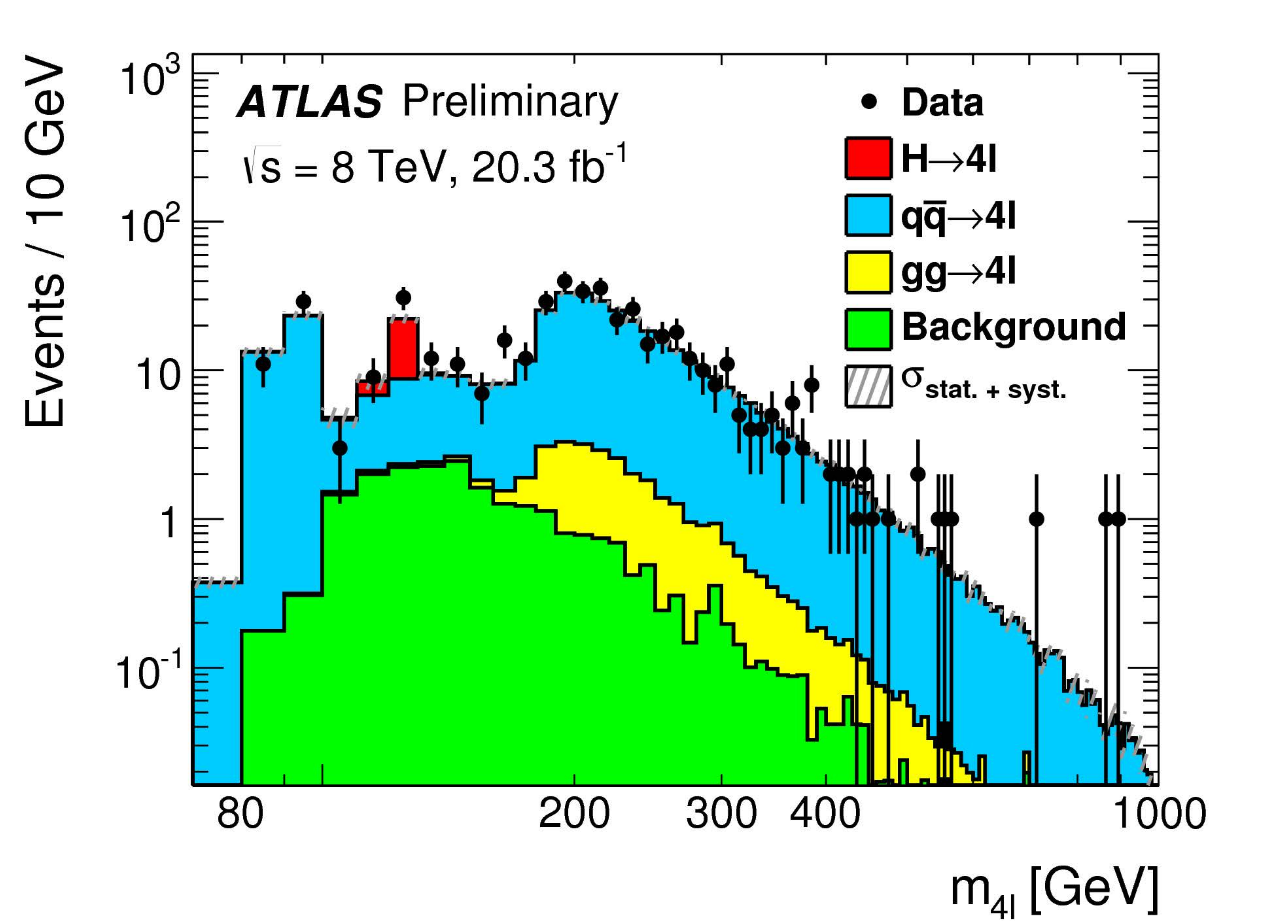}
\caption{Left: Measured and predicted $W^+W^-$ production cross sections at 7 and 8 TeV. Right: measured and predicted $m_{4\ell}$ distributions at 8 TeV.}
\label{fig:tgc}
\end{figure}

Both ATLAS and CMS measured the $pp \rightarrow W^+ W^- \rightarrow \ell\nu\ell'\nu'$ cross section using 7 TeV  and partial 8 TeV data~\cite{atlas_ww_7TeV}~\cite{cms_ww_7TeV}~\cite{cms_ww_8TeV_partial}. 
The measured fiducial cross sections are found to be $10-20$\% higher than the corresponding QCD NLO predictions. 
The question is whether this is due to missing higher-order corrections in theoretical predictions, upward fluctuations in data, or effects of new physics.  
Theorists recently performed QCD NNLO calculations and found the cross sections increased by about 8\% at $7-8$ TeV~\cite{ww_theory}. 
The NNLO predictions are found to be consistent with new measurements performed by both collaborations using the full 8 TeV 
dataset~\cite{atlas_ww_8TeV}~\cite{cms_ww_8TeV}. Figure ~\ref{fig:tgc} (left) shows the comparison of the measured and predicted cross sections at 7 and 8 TeV. 
To reduce the SM $t\bar{t}$ background, ATLAS vetoed events with any reconstructed jets, while 
CMS vetoed events with more than one reconstructed jet. In addition, CMS also vetoed events with any reconstructed $b-$jets.
To further check the modelling of the $p_T$ of the $WW$ system~\cite{ww_pT}, CMS measured the $WW+1$ jet cross section with the jet $p_T$ threshold set to 20, 25 
and 30 GeV. The measured fiducial cross sections for different jet $p_T$ thresholds are also found to be consistent with theoretical predictions~\cite{cms_ww_8TeV}. 

$W\gamma$ and $Z\gamma$ inclusive and differential cross section measurements performed by both collaborations using 7 TeV data~\cite{atlas_wzgamma_7TeV}~\cite{cms_wzgamma_7TeV} indicate that 
QCD NLO calculations are not accurate enough since worse data-theory agreement happened for events with high $p_T$ photons or large jet multiplicity.  
Theorists recently performed QCD NNLO calculations for these two processes~\cite{theory_wzgamma}. Better agreement is observed between predicted and measured 
cross sections for previous measurements and new CMS $Z\gamma$ measurement using the full 8 TeV dataset~\cite{cms_zgamma}.

ATLAS studied four lepton production in a mass range from 80 to 1000 GeV at 8 TeV~\cite{atlas_zz4l}. For $m_{4\ell}$ 
close to 90 GeV, it is mainly due to $q\bar{q} \rightarrow Z \rightarrow \ell\ell Z^{(*)} \rightarrow 4\ell$; for $m_{4\ell}$ around 125 GeV, it is mainly 
due to Higgs production through gluon-fusion $gg \rightarrow H \rightarrow ZZ^* \rightarrow 4\ell$; while for $m_{4\ell}$ above 180 GeV, it is mainly 
due to $t-$channel production of $q\bar{q} \rightarrow Z^{(*)}Z^{(*)} \rightarrow 4\ell$, non-resonant production through quark-box diagram of $gg \rightarrow Z^{(*)}Z^{(*)} \rightarrow 4\ell$, 
or through $s-$channel production of an off-shell Higgs boson. The measured differential cross section 
($d\sigma/dm_{4\ell}$) is found to be consistent with theoretical predictions with QCD NNLO and electroweak NLO calculations used for on-shell Higgs and 
$q\bar{q}$ production together with QCD LO calculations used for non-resonant $gg-$induced production.
 Figure ~\ref{fig:tgc} (right) shows the measured and predicted $m_{4\ell}$ distributions at 8 TeV. 
 Events with $m_{4\ell}>180$ GeV are used to extract the 
$gg-$induced signal strength with respect to the LO $gg$ prediction. The strength is found to be $2.4 \pm 1.0$ (stat) 
$\pm 0.5$ (syst) $\pm 0.8$ (theory).

CMS measured the inclusive cross section for the $pp \rightarrow ZZ \rightarrow \ell\ell\nu\nu$ process~\cite{cms_zzllvv}. This channel has a higher branching ratio 
compared to the $pp \rightarrow ZZ \rightarrow 4\ell$ channel but expects to have larger SM $Z(\rightarrow \ell\ell)+$ jets backgrounds. 
CMS chose to cut on the reduced MET and  MET balance variables. The MET vector is decomposed to 
the parallel and perpendicular directions along the dilepton $p_T$ axis. 
The reduced MET projection along direction $i$ is defined as $-p_T^{\ell\ell, i}- R^i$ where ${p_T^{\ell\ell}}$ is the $p_T$ of the dilepton system. 
The hadronic recoil vector ${\bf R}$ is calculated using clustered or unclustered jets in the event and the one with a minimal 
absolute value of that reduced MET component is used. 
The MET balance is defined as MET$/p_T^{\ell\ell}$. A cut on the reduced MET $>65$ GeV and $0.4<$MET balance$<1.8$ significantly reduced the $Z+$ jets 
background. The measured inclusive cross section is found to be consistent with the prediction at both 7 and 8 TeV.  

CMS measured the electroweak production of a single $W$ boson~\cite{cms_w_vbs}. The final state has one lepton, large MET, and two forward jets with large invariant mass 
and large pseudorapidity separation. The dominant background comes from the QCD production of $W+$ 2 jets and is estimated using MC-simulated 
events. The scale factor applied on the MC estimation is extracted from a template fit to the lower side of the BDT distribution for events with $M_{jj}>260$ GeV, 
which is dominated by QCD $W+$ 2 jets production. The $m_{jj}$ criteria is raised to 1000 GeV in the fiducial region and the signal contribution is obtained 
from a maximum-likelihood fit to the $m_{jj}$ distribution. The measured cross section is found to be consistent with the prediction. 

\section{Processes with Quartic Gauge Couplings}
There are two different ways to study QGC processes: one is through the triboson production where we have three bosons in the final state, and the 
other is through vector boson scattering/fusion processes where each colliding quark radiates a gauge boson and these two gauge bosons interact 
with each other and produce two bosons and two forward jets in the final state. QGC processes are often difficult to observe due to their relatively 
low production cross sections. In addition, we need to remove contributions from processes involving TGC vertices, fermion-mediated or Higgs-mediated 
vertices that produce the same final state. 

Since massless Goldstone bosons are absorbed by massless gauge bosons to become their longitudinal components, it is important to study the 
longitudinal scattering of two heavy gauge bosons ($V_LV_L \rightarrow V_LV_L$ with $V=W, Z$). In the SM, without Higgs-related diagrams, 
the cross section of VBS processes increases as a function of the center-of-mass energy of the diboson system and violates unitarity at 
above $\sim 1$ TeV~\cite{vbs_theory}. The introduction of a SM Higgs boson is the most economic way to restore the unitarity, however it is not the only way. It is possible that 
we have a Higgs boson with (non-)SM $HVV$ couplings and a (or several) heavy resonance(s) that restores unitarity for the whole energy regime.

\begin{figure}[htb]
\centering
\includegraphics[height=2.5in]{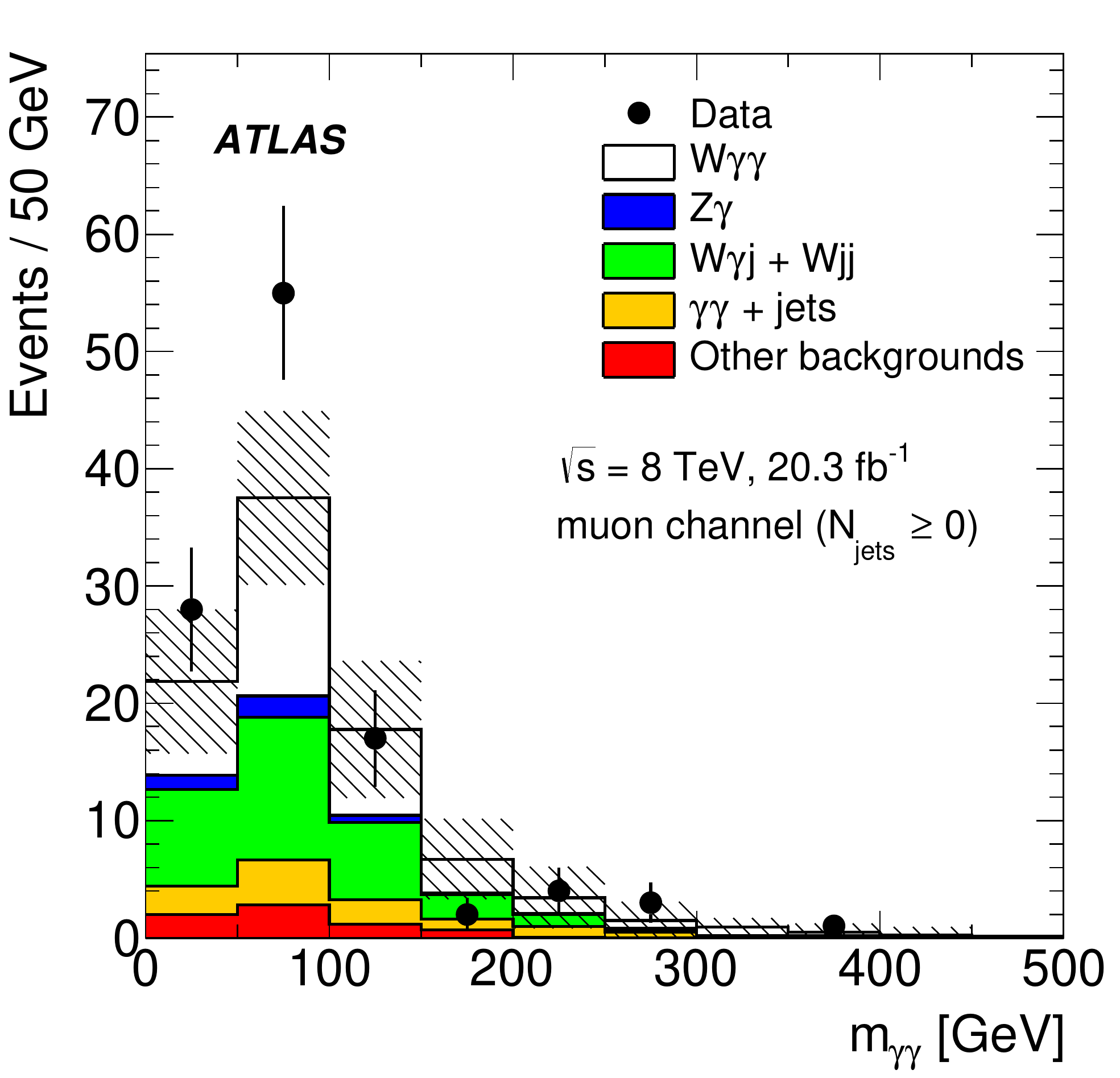}
\includegraphics[height=2.4in]{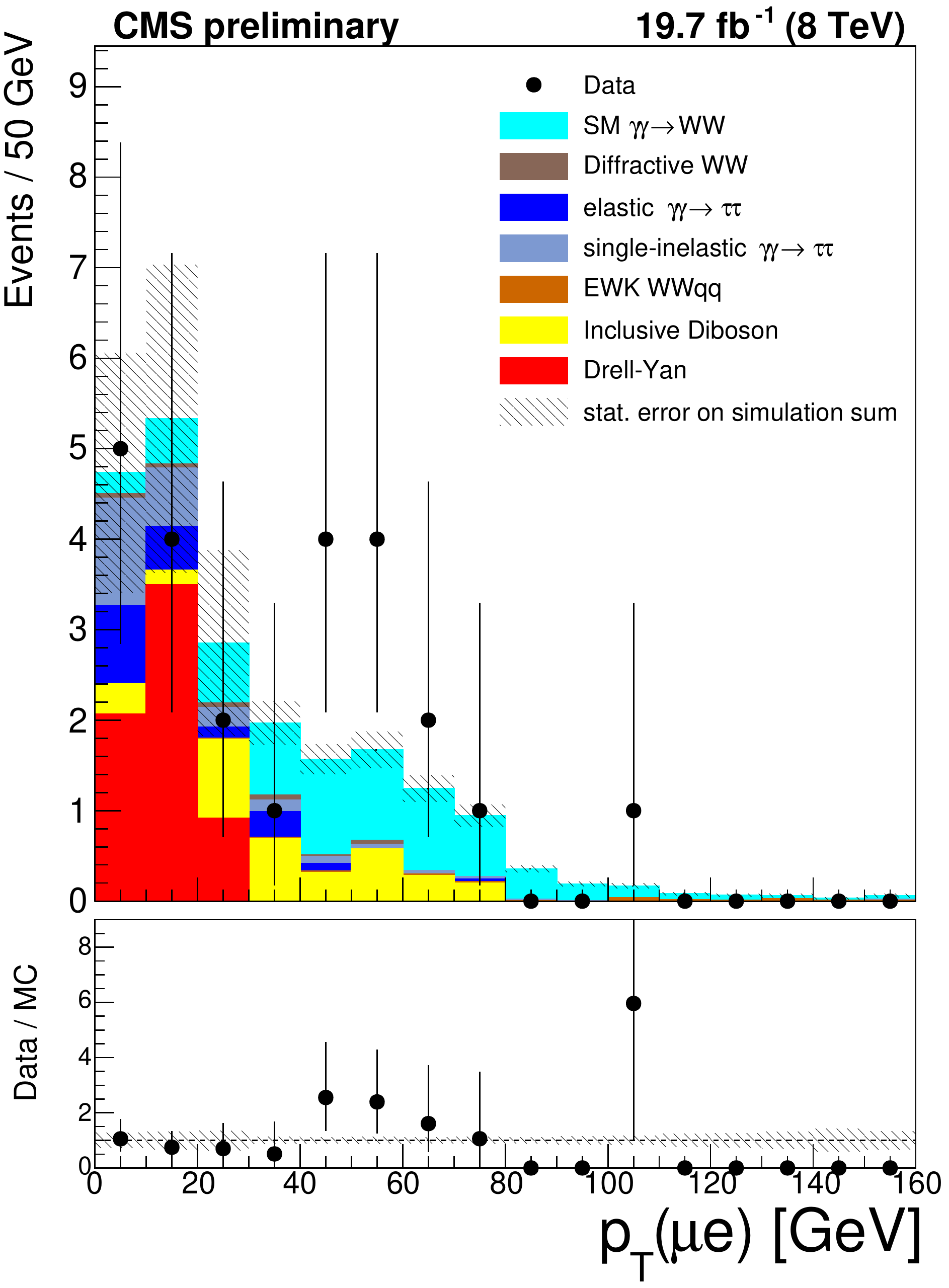}
\caption{Left: measured and predicted $m_{\gamma\gamma}$ distributions in the ATLAS $W\gamma\gamma$ analysis. Right: measured and predicted $p_T(e\mu)$ distribution in the CMS $\gamma \gamma \rightarrow WW$ analysis. }
\label{fig:qgc}
\end{figure}

Triboson $W\gamma\gamma$ process has the largest cross section among all triboson processes. ATLAS provided the first evidence for this process using 
the whole data collected at 8 TeV~\cite{atlas_wgammagamma}. Events are divided into inclusive ($N_{jet} \ge 0$) and exclusive ($N_{jet}=0$) categories. The dominant background comes from 
processes where jets are misidentified as photons or leptons. These backgrounds are estimated from a 2-dimensional template fit to the two photon 
isolation energy distribution (for the jet-faked background) or the lepton isolation and MET distribution (for the lepton-faked background). The significance
compared to the prediction without the $W\gamma\gamma$ process is 3 $\sigma$ and thus provides the first evidence for a triboson process at the LHC. 
Figure ~\ref{fig:qgc} (left) shows the measured and predicted $m_{\gamma\gamma}$ distributions. 

CMS searched for the $WV\gamma$ production with $W \rightarrow \ell\nu$ and $V \rightarrow jj$~\cite{cms_wvgamma}. Events are selected with an isolated lepton, an isolated photon, 
large MET, and two jets that have an invariant mass close to the $W$ and $Z$ boson mass. The dominant background comes from the $W\gamma+$jets process 
and its contribution is estimated using MC-simulated events. No evidence has been found for this processand upper limits on the production cross section is set. 

CMS searched for $\gamma \gamma \rightarrow WW$ process using $pp \rightarrow W^+ W^- p^{(*)} p^{(*)}$  events~\cite{cms_gammagammaww}, where each proton radiates a 
photon and the final state protons have very forward pseudorapidities and are not detected. To reduce the Drell-Yan background, only the $e\mu$ final state 
is used. Signal events tend to have a high $p_T$ isolated opposite-sign $e\mu$ pair with large $p_T(e\mu)$ and $m_{e\mu}$. Besides the 
electron and muon tracks, there are no other tracks associated with the primary collision vertex. The data is found to be 3.6 $\sigma$ above 
the background-only hypothesis. Figure ~\ref{fig:qgc} (right) shows the measured and predicted $p_T(e\mu)$ distributions. 
The dominant background for this process comes from exclusive $\gamma \gamma \rightarrow \ell^+ \ell^-$ production, and ATLAS performed a measurement in 
both electron and muon channels at 7 TeV~\cite{atlas_gammagammall}. The measured cross sections are found to be consistent with theoretical predictions 
with proton absorptive effects due to the finite size of the proton taken into account.

Both ATLAS and CMS collaborations searched for the same-sign $WW$ VBS process with two same-sign dilepton, large MET and two forward jets that 
has a large invariant mass and are well separated in pseudorapidity~\cite{atlas_ssww}~\cite{cms_ssww}. ATLAS measured the cross section in an inclusive signal region and also a region dominated 
by electroweak production. The significance is found to be 4.5$\sigma$ and 3.6$\sigma$ above the background-only hypothesis for these two regions, respectively.
CMS also performed a measurement of the $WZ+2$ jets cross section.

\section{Conclusions}
Table~\ref{tab:xsection} shows measured and predicted cross sections for TGC and QGC processes from both collaborations. 
Good agreement is observed for all processes. 

\begin{table}[t]
\begin{center}
\begin{tabular}{l|cc}   
\hline \hline
Process (COM energy) & $\sigma_{\mbox{measured}}$  &  $\sigma_{\mbox{predicted}}$  \\ \hline
 ATLAS  &     &             \\
$pp \rightarrow W^+ W^-$ (8) &  $71.4 \pm 1.2$(stat)$^{+5.0}_{-4.4}$(syst)$^{+2.2}_{-2.1}$(lumi) pb   &     $63.2 \pm 2.0$  pb\\
$pp \rightarrow 4\ell$ (8) &  $73 \pm 4$(stat)$\pm 4$(syst)$\pm 2$(lumi) fb     &     $65 \pm 4$ fb \\
$pp \rightarrow W\gamma\gamma$ ($N_{jet} \ge 0$) (8) & $6.1^{+1.1}_{-1.0}$(stat)$\pm 1.2$(syst)$\pm 0.2$(lumi) fb   &     $2.90 \pm 0.16$  fb\\
$pp \rightarrow W\gamma\gamma$ ($N_{jet} > 0$) (8) & $2.9^{+0.8}_{-0.7}$(stat)$^{+1.0}_{-0.9}$(syst)$\pm 0.1$(lumi) fb   &     $1.88 \pm 0.20$  fb\\
$pp \rightarrow W^\pm W^\pm jj$ (8) & $1.3 \pm 0.4$(stat)$\pm 0.2$(syst) fb   &     $0.968 \pm 0.06$  fb\\
$\gamma \gamma \rightarrow \ell^+ \ell^-$ ($\ell=e$) (7) & $0.428 \pm 0.035$(stat)$\pm 0.018$(syst) pb & $0.398 \pm 0.007$ pb \\
$\gamma \gamma \rightarrow \ell^+ \ell^-$ ($\ell=\mu$) (7) & $0.628 \pm 0.032$(stat)$\pm 0.021$(syst) pb & $0.638 \pm 0.011$ pb \\
 CMS  &     &                \\
$pp \rightarrow W^+ W^-$ (8) &  $60.1 \pm 0.9$(stat)$\pm 4.5$(syst)$\pm 1.6$(lumi) pb   &     $59.8^{+1.3}_{-1.1}$  pb\\
$ZZ \rightarrow \ell\ell\nu\nu$ (7) &  $67^{+20}_{-18}$(stat)$^{+18}_{-14}$(syst)$\pm 2$(lumi) fb     &     $79^{+4}_{-3}$ fb     \\
$ZZ \rightarrow \ell\ell\nu\nu$ (8) &  $88^{+11}_{-10}$(stat)$^{+24}_{-18}$(syst)$\pm 4$(lumi) fb     &     $97^{+4}_{-3}$ fb     \\
EWK $W+$ 2 jets (8) & $0.42 \pm 0.04$(stat)$\pm 0.09$(syst)$\pm 0.01$(lumi) pb    &     $0.50 \pm 0.03$ pb  \\
$pp \rightarrow p^{(*)} \mu^\pm e^{\mp} p^{(*)}$ (8) & $12.3 ^{+5.5}_{-4.4}$ fb   & $6.9 \pm 0.6$ fb \\
$pp \rightarrow WV\gamma$ (8) & $<311$ fb   & $91.6 \pm 21.7$ fb \\
$pp \rightarrow W^\pm W^\pm jj$ (8) & $4.0^{+2.4}_{-2.0}$(stat)$^{+1.1}_{-1.0}$(syst) fb   &     $5.8 \pm 1.2$  fb\\
\hline \hline
\end{tabular}
\caption{Measured and predicted cross sections for TGC and QGC processes from both collaborations at different center-of-mass (COM) energies. The cross sections are different for 
the same process due to different phase spaces used.}
\label{tab:xsection}
\end{center}
\end{table}

\begin{figure}[htb]
\centering
\includegraphics[height=2.in]{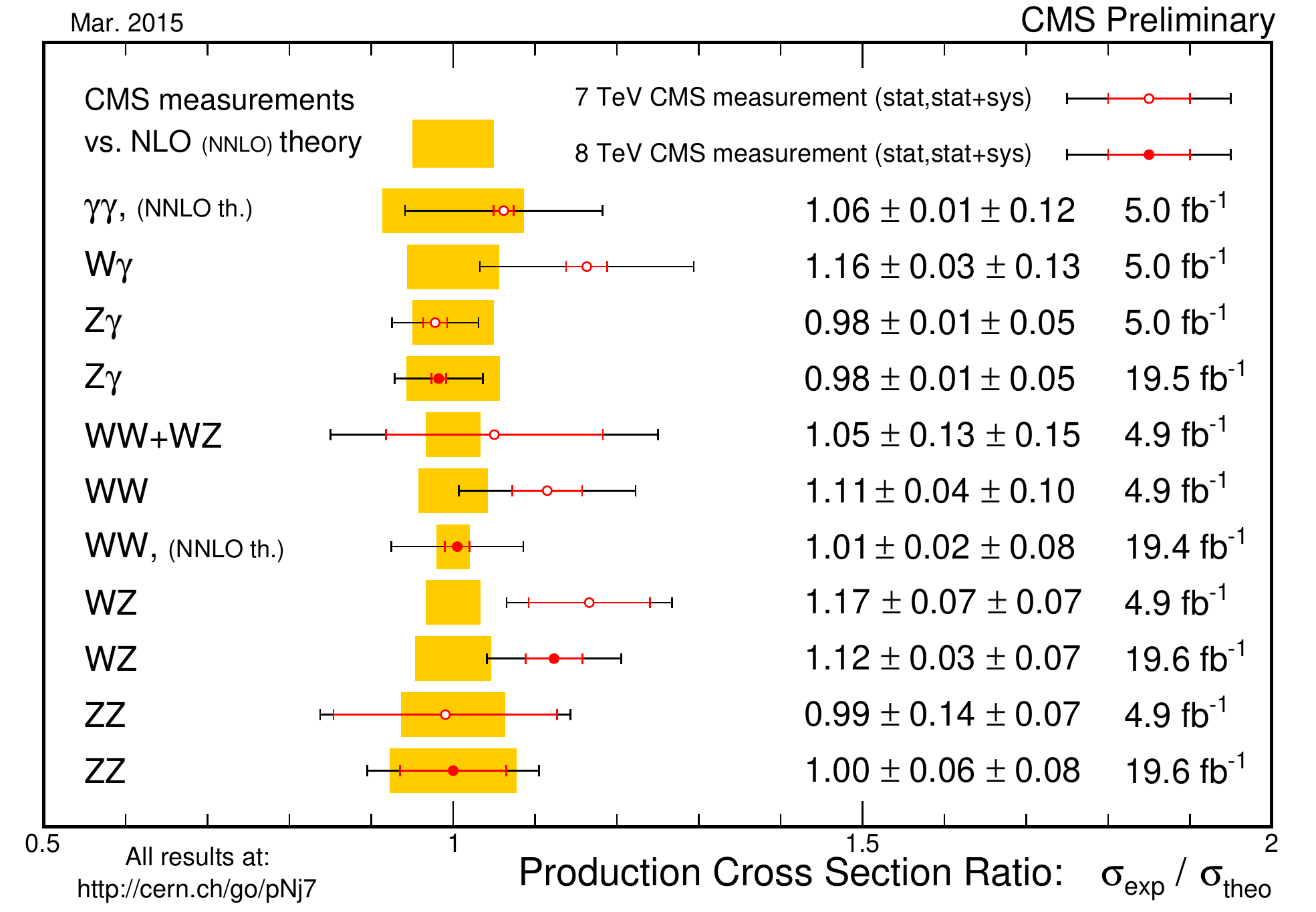}
\includegraphics[height=2.1in]{SMSummary_ATLAS}
\caption{Summary of several Standard Model total and fiducial production cross section measurements, corrected for leptonic 
branching fractions, compared to the corresponding theoretical expectations. All theoretical expectations were calculated at NLO or higher. }
\label{fig:xsection_summary}
\end{figure}

Figure~\ref{fig:xsection_summary} (left) shows the ratio of the measured and predicted cross sections for diboson processes measured by the CMS 
collaboration, good agreement is observed for all processes here. The experimental uncertainties are around a few percent for most processes. 
Figure~\ref{fig:xsection_summary} (right) shows the measured and predicted cross sections for all SM processes performed at ATLAS. The same-sign $WW$ VBS 
process has the lowest cross section that is around 1 fb. 

The direct Higgs measurement at the LHC provides a critical input to the global electroweak fit. It is important to improve $m_W$ and $\sin^2 \theta^{\ell}_{eff}$  measurements at the LHC 
to further check the SM consistency. A better understanding of PDF uncertainty is critical to reduce systematic uncertainties on these two measurements. 

With large production cross sections for diboson processes, we are working toward precise measurements of diboson processes at the LHC. 
Experimental precisions for some processes already matched theoretical predictions. Higher-order corrections beyond the QCD NLO and electroweak LO 
calculations are needed for the comparison. 

We started to observe processes with production cross sections close to 1 fb and provided first evidences for $W\gamma\gamma$ triboson process. In 
addition, we also provided first evidences for $\gamma\gamma \rightarrow W^+ W^-$ and same-sign $WW$ VBS processes. With a larger dataset expected 
at Run 2, we will observe more triboson and VBS processes and perform more precise measurements. It is important to check the high-mass behavior of 
VBS processes to gain a better understanding of the electroweak symmetry breaking mechanism.

\Acknowledgments
I am grateful to the DPF 2015 organization committee for inviting me to give this overview talk.

\end{document}